\documentclass[prd,onecolumn,preprintnumbers,nofootinbib,showpacs,showkeys]{revtex4}
\usepackage{latexsym,graphicx,amssymb,amsmath,mathrsfs}
\usepackage{setspace,bm} 
\DeclareSymbolFont{bbold}{U}{bbold}{m}{n}
\DeclareSymbolFontAlphabet{\mathbbold}{bbold}

\newcommand{\be}{\begin{equation}}      
\newcommand{\ee}{\end{equation}}      
\newcommand{\bea}{\begin{eqnarray}}      
\newcommand{\eea}{\end{eqnarray}}    
     
\newcommand{\rt}[1]{{}}      
  
\newcommand{\ifff}{\,\textrm{if}\,}  
\newcommand{\elsee}{\,\textrm{else}\,}  
\newcommand{\Tr}{\,\textrm{Tr}\,}
\newcommand{\diag}{\,\textrm{diag}\,}

\makeatletter
\renewcommand\appendix{\par
\setcounter{section}{0}%
\setcounter{subsection}{0}%
\gdef\thesection{\appendixname\space\@Alph\c@section}}

\long\def\unmarkedfootnote#1{{\long\def\@makefntext##1{##1}\footnotetext{#1}}}
\makeatother

\begin{document} 

\title{Chiral symmetry breaking patterns in the $U_L(n)\times U_R(n)$ meson model} 
\preprint{RIKEN-QHP-60}
\author{G. Fej\H{o}s}
\email{fejos@riken.jp}
\affiliation{Theoretical Research Division, Nishina Center, RIKEN, Wako 351-0198, Japan}

\begin{abstract}
{Chiral symmetry breaking patterns are investigated in the $U_L(n)\times U_R(n)$ meson model. It is shown that new classes of minima of the effective potential belonging to the center of the Lie algebra exist for arbitrary flavor number $n$. The true ground state of the system is searched nonperturbatively and although multiple local minima of the effective potential may exist, it is argued that in regions of the parameter space applicable for the strong interaction, strictly a $U_L(n)\times U_R(n)\longrightarrow U_V(n)$ spontaneous symmetry breaking is possible. The reason behind this is the existence of a discrete subset of axial symmetries, which connects various $U_V(n)$ symmetric vacua of the theory. The results are in agreement with the Vafa-Witten theorem of QCD, illustrating that it remains valid, even without gauge fields, for an effective model of the strong interaction.}
\end{abstract}

\pacs{11.30.Qc, 11.30.Rd}
\keywords{Chiral symmetry breaking}  
\maketitle

\section{Introduction}

Quantum chromodynamics (QCD) displays an approximate global $U_L(3)\times U_R(3)$ chiral symmetry. The observed mesonic spectra show that this symmetry is broken spontaneously in the ground state of QCD matter. Around the temperature of the scale of QCD ($\Lambda_{QCD}$), strongly interacting matter undergoes a chiral symmetry restoration, accompanied by the quark deconfinement transition \cite{fukushima11b}. Due to difficulties of solving QCD at low energies, effective models built upon the approximate chiral symmetry can give reasonable results regarding the properties of the chiral transition and therefore the phase diagram of the strongly interacting matter. The $n$-flavored matrix model displaying a $U_L(n) \times U_R(n)$ symmetry \cite{gellmann60} gives decent results on the mesonic spectra and phase diagram \cite{lenaghan00,roder03,herpay05} of QCD, and can be extended with quark degrees of freedom (quark-meson model) \cite{bilic99,kovacs07} and/or the Polyakov loop \cite{kahara08,mao10,marko10,schaefer10,herbst11,tiwari12}, tetraquark excitations \cite{fariborz11,mukherjee12}, and vector mesons \cite{parganlija13}. Although none of these models can be solved exactly either, they provide a simpler theoretical framework than QCD to describe the behavior of the strongly interacting matter at finite temperature and density. The $U_A(1)$ anomaly of QCD can be easily included into the model through 't Hooft's determinant term \cite{thooft76}; however, it is argued that close to the critical point the instanton density causing the anomaly is suppressed exponentially \cite{schafer98}. Furthermore, recently new signs of the restoration of the $U_A(1)$ factor around the critical temperature appeared in the literature for $n=2$ \cite{aoki12}; therefore, close to the phase transition the anomaly is not expected to play an important role and the pure $U_L(n)\times U_R(n)$ symmetric model might have phenomenological relevance.

It has been known for a long time that there is no infrared stable fixed point of the renormalization group flow of the $U_L(n)\times U_R(n)$ model in $4-\epsilon$ ($\epsilon>0$) dimensions \cite{pisarski84}: $\epsilon$ expansion shows that if $n\geq 2$, even the existing nontrivial $O(2n^2)$-like fixed point becomes unstable in the direction of the quartic coupling. The nonexistence of an infrared stable fixed point indirectly shows a transition, which (if exists) is fluctuation induced, and of first order. This has been confirmed numerically for the $n=2$ case with the use of the functional renormalization group (FRG) approach \cite{berges97,berges97b,fukushima11}, and expected to remain true at $n>2$.

Analysis of the fixed point structure does not tell anything about the symmetry breaking pattern; however, studies of the $U_L(n)\times U_R(n)$ model in the past assumed {\it a priori} that the symmetry is broken to $U_V(n)$. This means that although due to explicit symmetry breaking terms several nonzero condensate components might appear \cite{lenaghan00,herpay05,fariborz11,mukherjee12}, spontaneous symmetry breaking and its recovery has been observed in the piece proportional to the unit matrix \cite{herpay05}. Besides \cite{fejos12}, there were no attempts to search for spontaneous symmetry breaking in other order parameters. Presumably, the reason behind this is the Vafa-Witten theorem \cite{vafa84}, which states that in vectorlike gauge theories with vanishing $\theta$ angles (e.g. QCD) and nonzero but equal fermion masses, global vector symmetries can not be broken spontaneously. Since the $U_L(n)\times U_R(n)$ symmetric theory is an effective model of QCD, it is plausible to accept that the original statement remains true; however, it should not be considered correct without further exploration. This is especially true for vanishing quark masses (i.e. no explicit breaking of chiral symmetry), for which the theorem only states that if a vector-symmetry breaking state existed, its energy could not be lower than of the symmetry conserving one. The situation gets puzzling, if we accept the results of Ref. \cite{fejos11}, where (in the large-$n$ limit) a new local minimum of the energy functional realizing a symmetry breaking pattern $U_L(n)\times U_R(n) \longrightarrow U_V(n-1)$ was found, from which the system can undergo a first order transition with appropriately chosen model parameters \cite{fejos12}. 

Furthermore, in \cite{nardi11,espinosa12} it was conjectured that Yukawa couplings of the Standard Model might be described by the spontaneous breaking of a hypothetic $SU(3)\times SU(3)$ symmetry, realizing yet another ground state which breaks the symmetry to $SU(2)\times SU(2)\times U(1)$. This solution has already been reported in \cite{paterson81} for a generic number of flavors. The existence of these different vacua in the $U_L(n)\times U_R(n)$ model might be in apparent contradiction with the Vafa-Witten theorem; therefore, it is an open question whether these minima are not physically applicable for some reason, or the applied approximations are inconsistent.

The goal of this paper is to answer the aforementioned questions and to map the vacuum structure in an extended space of the condensate space, searching for new local minima of the effective potential for arbitrary flavor numbers. The paper is organized in a pedagogical way and goes from the most simple cases to a general analysis. In Sec. II, we introduce the model and the symmetry breaking we would like to study in the first place. We list the group invariants and discuss the possibility of breaking of the diagonal $U_V(n)$ symmetry. In Sec. III, we solve the field equations at classical level, and show the energy properties with the corresponding spectra of the found solutions. In Sec. IV we include quantum fluctuations using FRG formalism. We start with the $n=2$ case in the so-called local potential approximation (LPA), then generalize the arguments for the exact FRG formulation for $n\geq 2$. It is followed by an analysis of patterns described by more general condensates, and a discussion of obtaining all the $U_V(n)$ symmetric ground states. Conclusions can be found in Sec. V.

\section{The model and spontaneous symmetry breaking}

The dynamical variable of the $U_L(n)\times U_R(n)$ scalar theory is a matrix field $M$, as an element of the $U(n)$ Lie algebra. It can be parametrized as
\bea
M=(s^a+i\pi^a)T^a, \qquad (a=0...,n^2-1),
\eea
where $s^a$ and $\pi^a$ are the scalar and pseudoscalar fields, respectively, with $T^a$ being the group generators satisfying $\Tr (T^a T^b)=\delta^{ab}/2$. The Lagrangian of the model is
\bea
{\cal L}=\partial_\mu M \partial^{\mu} M^\dagger - m^2 \Tr (MM^\dagger) -\frac{g_1}{n^2}[\Tr (MM^\dagger)]^2- \frac{g_2}{n}\Tr (MM^\dagger M M^\dagger),
\label{Eq:lag}
\eea
where $m^2<0$. In order to have a classical potential bounded from below, we need to require
\bea
g_1[\Tr(MM^\dagger)]^2+ng_2\Tr(MM^\dagger MM^\dagger)>0
\label{Eq:bound}
\eea
for arbitrary field configurations. It is clear that dilation $M\longrightarrow \lambda M (\lambda > 0)$ leaves this condition invariant; therefore, we can choose $\Tr (MM^\dagger)=1$. Furthermore, we can always diagonalize $M$ in a way that the potential does not change, therefore without the loss of generality, we can assume that $M$ is diagonal and $\Tr (MM^\dagger)=1$. Then we have
\bea
g_1+ng_2\Tr(MM^\dagger MM^\dagger)>0,
\eea
with $M$ being diagonal. Using a Lagrange multiplier, it is easy to show that among diagonal $M$ matrices with $\Tr (MM^\dagger)=1$,
\bea
1 \geq \Tr(MM^\dagger MM^\dagger) \geq 1/n;
\eea
therefore, the stability region of the potential is determined by $g_1+g_2>0$ and $g_1+ng_2>0$. If $g_2>0$, then $g_1+g_2>0$ is the stronger condition; if $g_2<0$, then $g_1+ng_2>0$. Since without $U_A(1)$ anomaly, $g_2$ is proportional to the kaon-pion mass square difference: $g_2\sim m_K^2-m_\pi^2$ \cite{lenaghan00}; in this paper we restrict ourselves to $g_2>0$ (and therefore $g_1+g_2>0$).

It is clear that, (\ref{Eq:lag}) is invariant under the symmetry transformation
\bea
M \longrightarrow U_R M U_L^\dagger,
\eea
$U_L$ and $U_R$ are independent $U(n)$ matrices realizing the chiral symmetry with parameters $\theta_R^a$ and $\theta_L^a$. These transformations can be reformulated in terms of vector and axial-vector symmetries as:
\bea
M \longrightarrow V M V^\dagger, \hspace{1.5cm} M \longrightarrow A^\dagger M A^\dagger,
\eea
where $V$ and $A$ are also unitary matrices with parameters $\theta^a_{V,A}=(\theta^a_R \pm \theta^a_L)/2$, respectively. Even though in QCD the axial symmetry is broken anomalously at quantum level, in this paper we do not investigate quantitatively the role of this anomaly, nor any explicit symmetry breaking. Following \cite{fejos11}, first we are interested in the possibility of the symmetry breaking pattern
\bea
\label{Eq:symbre}
<\!\!M\!\!>=v_0 T^0+v_8 T^8,
\eea
which means that we set all other expectation values in advance to zero, and investigate only on the $v_0-v_8$ plane. We will come to the question of general condensates later. In (\ref{Eq:symbre}) $T^0$ is the generator of the $U(1)$ subgroup of $U(n)$ and $T^8$ refers to the longest diagonal generator, regardless of the actual number of flavors (see also Appendix A). This notation is just an adaption from the phenomenologically most important $n=3$ case. Note that, pattern (\ref{Eq:symbre}) does not apply for $n=2$, since in this case there are only four generators. Besides $T^0$, only one of them is diagonal: $T^3$, the third Pauli matrix. Formulas given below can be applied to $n=2$ by considering $T^3$ instead of $T^8$, i.e. substitute index ``3'' instead of ``8'' everywhere.

In terms of the coefficient fields $s^a$ and $\pi^a$, (\ref{Eq:lag}) turns into
\bea
{\cal L}=\frac12 \partial_\mu s^a \partial^\mu s^a+\frac12 \partial_\mu \pi^a \partial^\mu \pi^a - V(s^a,\pi^a),
\eea
where
\bea
V = m^2 I_1(s^a,\pi^a)+\frac{g_1}{n^2}I_1^2(s^a,\pi^a)+\frac{g_2}{n}I_2(s^a,\pi^a),
\label{Eq:V}
\eea
with the use of notations
\begin{subequations}
\bea
I_1(s^a,\pi^a)&\equiv&\Tr(MM^\dagger)=\frac12(s^as^a+\pi^a\pi^a), \\
I_2(s^a,\pi^a)&\equiv&\Tr(MM^\dagger M M^\dagger)=\frac18d^{abc}d^{aef}(s^bs^c+\pi^b\pi^c)(s^es^f+\pi^e\pi^f)+\frac12f^{abe}f^{acf}s^bs^c\pi^e\pi^f.
\eea
\end{subequations}
Here $d^{abc}$ and $f^{abc}$ are the totally symmetric and antisymmetric tensors of the $U(n)$ group, respectively, see details in Appendix A. We will need the invariants and their derivatives in a background of $v_0, v_8$. The invariants are:
\bea
I_1\big|_{v_0,v_8}=\frac12(v_0^2+v_8^2), \qquad I_2\big|_{v_0,v_8}=\frac{1}{4n}(v_0^4+v_8^4)+\frac18(d^{888})^2 v_8^4+\frac{3}{2n}v_0^2v_8^2+\frac{1}{\sqrt{2n}}d^{888}v_8^3v_0.
\label{Eq:inv}
\eea
The needed structure constants are listed in Appendix A; the derivatives of the invariants can be found in Appendix B. The tree level mass matrices are defined as follows:
\bea
M_{ij}^{2(s)}=\frac{\partial^2 V}{\partial s^i \partial s^j}\bigg|_{v_0,v_8}, \qquad M_{ij}^{2(\pi)}=\frac{\partial^2 V}{\partial \pi^i \partial \pi^j}\bigg|_{v_0,v_8}.
\eea
At classical level these can be written as
\begin{subequations}
\bea
\label{Eq:mass_s}
M_{ij}^{2(s)}&=&\left(m^2+\frac{2g_1}{n^2}I_1|_{v_0,v_8}\right) \frac{\partial^2 I_1}{\partial s_i \partial s_j}\bigg|_{v_0,v_8}+\frac{g_2}{n}\frac{\partial^2 I_2}{\partial s_i \partial s_j}\bigg|_{v_0,v_8}+\frac{2g_1}{n^2}\frac{\partial I_1}{\partial s_i}\bigg|_{v_0,v_8}\frac{\partial I_1}{\partial s_j}\bigg|_{v_0,v_8},\\
\label{Eq:mass_p}
M_{ij}^{2(\pi)}&=&\left(m^2+\frac{2g_1}{n^2}I_1|_{v_0,v_8}\right) \frac{\partial^2 I_1}{\partial \pi_i \partial \pi_j}\bigg|_{v_0,v_8}+\frac{g_2}{n}\frac{\partial^2 I_2}{\partial \pi_i \partial \pi_j}\bigg|_{v_0,v_8}.
\eea
\end{subequations}
Checking the detailed expressions of (\ref{Eq:mass_s}) and (\ref{Eq:mass_p}) found in Appendix B, it can be seen that we have to diagonalize in the $0-8$ indices. This leads to the multiplet structure $1\oplus 1\oplus n(n-2)\oplus 2(n-1)$ in each sector. Note that, when $n=3$ these groups correspond to $\eta$, $\eta'$, three pions and four kaons in the pseudoscalar sector, and $\sigma$, $f_0$ singlets with three $a_0$'s and four $\kappa$'s in the scalar sector. In general, the spectrum may show the possibility of a $U_L(n)\times U_R(n) \longrightarrow U_V(n-1)$ symmetry breaking. We need to investigate if this indeed realizes in the minimum of the effective potential.

\section{New classical solutions of the field equations}
In this section we present the field equations for the condensates $v_0, v_8$ and search for all possible solutions of them. We note that the formulas below can be applied to $n=2$ as well, we just have to consider index 3 instead of 8 everywhere. Using (\ref{Eq:deriv1}), the field equations are
\begin{subequations}
\label{Eq:EoSgen}
\bea
0&=&\frac{\partial V}{\partial s^0}\bigg|_{v_0,v_8}=m^2v_0+\frac{g_1}{n^2}(v_0^2+v_8^2)v_0+\frac{g_2}{n^2}\left(v_0^3+3v_8^2v_0-\frac{n-2}{\sqrt{n-1}}v_8^3\right), \\
0&=&\frac{\partial V}{\partial s^8}\bigg|_{v_0,v_8}=m^2v_8+\frac{g_1}{n^2}(v_0^2+v_8^2)v_8+\frac{g_2}{n^2}\left(3v_0^2v_8-3\frac{n-2}{\sqrt{n-1}}v_8^2v_0+\frac{n^2-3n+3}{n-1}v_8^3\right).
\eea
\end{subequations}
Being cubic, these coupled equations can be solved analytically and we obtain nine solutions. Since the potential reflects $(v_0,v_8) \longleftrightarrow (-v_0,-v_8)$ symmetry, we have only five inequivalent solutions. Not considering the parity pairs, we get (see also Fig. 1):
\begin{subequations}
\label{Eq:eosfull}
\bea
\label{Eq:eosa}
v_0^{[I]}&=&0, \hspace{5cm} v_8^{[I]}=0, \\
\label{Eq:eosb}
v_0^{[II]}&=&\sqrt{\frac{-m^2n^2}{g_1+g_2}},\hspace{3.5cm} v_8^{[II]}=0, \\
\label{Eq:eosc}
v_0^{[III]}&=&(n-2)\sqrt{\frac{-m^2}{g_1+g_2}}, \hspace{2.3cm} v_8^{[III]}=2\sqrt{\frac{-m^2(n-1)}{g_1+g_2}}, \\
\label{Eq:eosd}
v_0^{[IV]}&=&(n-1)\sqrt{\frac{-m^2n}{g_1(n-1)+g_2n}}, \hspace{1.1cm} v_8^{[IV]}=\sqrt{\frac{-m^2n(n-1)}{g_1(n-1)+g_2n}}, \\
\label{Eq:eose}
v_0^{[V]}&=&\sqrt{\frac{-m^2}{g_1/n+g_2}}, \hspace{3.25cm} v_8^{[V]}=-\sqrt{\frac{-m^2(n-1)}{g_1/n+g_2}}.
\eea
\end{subequations}
We can immediately see that, for $n=2$, solutions (\ref{Eq:eosfull}) extended with parity pairs are symmetric under the interchange of the two condensates. This is due to the fact that at $n=2$, there is a $v_0 \longleftrightarrow v_8$ (or more appropriately $v_0 \longleftrightarrow v_3$) symmetry of the Lagrangian and of the field equations as well. In this case solution [V] turns out to be the reflection of solution [IV] with respect to the $v_8=0$ line, reflecting independent $v_0 \longleftrightarrow -v_0$, $v_3 \longleftrightarrow -v_3$ symmetries as well (this is not true for $n>2$).
\begin{figure}
\includegraphics[bb=50 50 554 770,angle=270,scale=0.46]{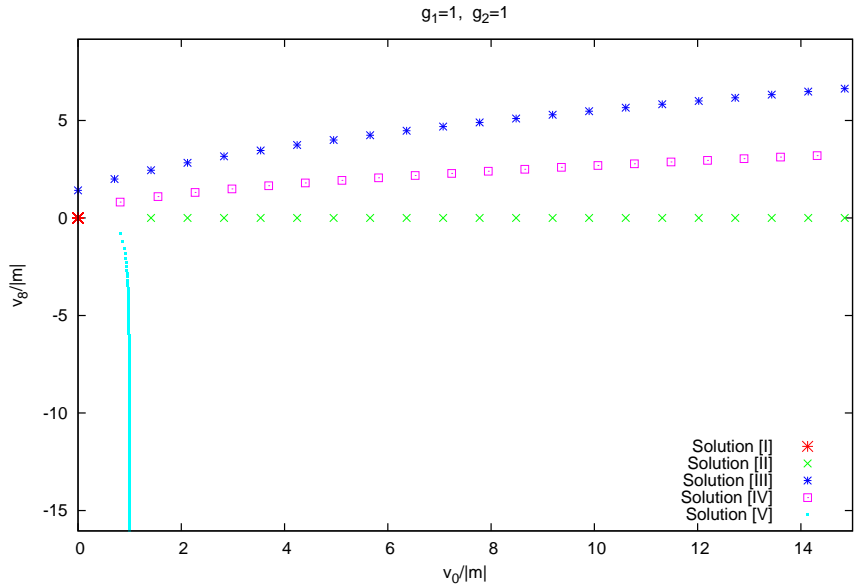}
\caption{Solutions of field equations (\ref{Eq:EoSgen}) for arbitrary $n\geq 2$. As $n$ increases, the solutions (except [I]) get away from the origin. Solution [I] is a local maximum, [II] and [III] are local minima, [IV] and [V] are saddle points.}
\end{figure}

Solution [V] shows $<\!\!M\!\!>\sim \diag(0,0,...,1)$, which has already been reported in \cite{paterson81} and particularly for $n=3$ in \cite{nardi11,espinosa12}, where it has been conjectured to be a ground state which may describe the hierarchy of Yukawa couplings in the Standard Model. We would like to mention however that, in the stability region which can be associated with QCD (i.e. $g_1+g_2>0$, $g_2>0$), this solution always describes a saddle point (this can be verified by checking the mass matrices [\ref{Eq:massmatrices}]), and therefore not a local minimum of the potential. In general, solutions [II] and [III] are local minima, [I] is a local maximum, and the others are saddle points. We note however that, if the couplings satisfied $g_2<0$ and $g_1+ng_2>0$, the only local minimum would be solution [V]. This region of the parameter space might be applicable for describing new interesting physical phenomena \cite{nardi11,espinosa12}, but since it can not have relevance in QCD, we do not investigate it further.

In the large-$n$ limit, solutions [II]-[III]-[IV] (after rescalings $v_0 \rightarrow nv_0$, $v_8 \rightarrow \sqrt{n}v_8$) exactly reproduce the three local extrema found in \cite{fejos11}. In this case solutions of $v_0$ become degenerate, and a double-well potential forms in the $v_8>0$ direction. The appearing new local minimum (solution [III]) was a candidate in \cite{fejos12} to describe the ground state and a first order transition of the system; see also Fig. 2. Our general analysis shows that this structure is although distorted at finite $n$, it still exists: a new local minimum (solution [III]) emerges for every each $n$ value.

The energies of the solutions are as follows:
\bea
V|_{v_0^{[I]},v_8^{[I]}}=0, \qquad V|_{v_0^{[II]},v_8^{[II]}}=-\frac{m^4n^2}{4(g_1+g_2)}, \qquad V|_{v_0^{[III]},v_8^{[III]}}=-\frac{m^4n^2}{4(g_1+g_2)}, \nonumber\\
V|_{v_0^{[IV]},v_8^{[IV]}}=-\frac{m^4n^2(n-1)}{4n(g_1+g_2)-4g_1}, \qquad V|_{v_0^{[V]},v_8^{[V]}}=-\frac{m^4n^2}{4(g_1+ng_2)},
\eea
and the following relations hold between them (using obvious shorthand notations):
\bea
V_{[II]}=V_{[III]}<V_{[IV]}\leq V_{[V]}<V_{[I]}.
\eea
Equality between $V_{[IV]}$ and $V_{[V]}$ is only possible for $n=2$, as it should. We see that having two minima with exactly the same energy is a general property for arbitrary $n$ (it is interesting to mention that, the energy over the number of degrees of freedom is even $n$ independent in these minima). This could make us suspicious, whether there is a symmetry forcing these minima to have the same amount of energy. Let us check the spectrum in these minima as well. First we calculate the masses which belong to [II]:
\begin{subequations}
\label{Eq:massII}
\bea
M^2_{\eta,[II]}&=&0, \hspace{1.7cm} M^2_{\eta',[II]}=0,  \hspace{1.7cm} M^2_{\pi,[II]}=0,  \hspace{1.7cm} M^2_{K,[II]}=-m^2\frac{2g_2}{g_1+g_2}, \\
M^2_{\sigma,[II]}&=&-m^2, \quad M^2_{f_0,[II]}=-m^2\frac{2g_2}{g_1+g_2}, \quad M^2_{a_0,[II]}=-m^2\frac{2g_2}{g_1+g_2}, \quad M^2_{\kappa,[II]}=0,
\eea
\end{subequations}
and then the ones corresponding to [III]:
\begin{subequations}
\label{Eq:massIII}
\bea
M^2_{\eta,[III]}&=&0, \hspace{1.7cm} M^2_{\eta',[III]}=0,  \hspace{1.7cm} M^2_{\pi,[III]}=0,  \hspace{1.7cm} M^2_{K,[III]}=0, \\
M^2_{\sigma,[III]}&=&-m^2, \quad M^2_{f_0,[III]}=-m^2\frac{2g_2}{g_1+g_2}, \quad M^2_{a_0,[III]}=-m^2\frac{2g_2}{g_1+g_2}, \quad M^2_{\kappa,[III]}=-m^2\frac{2g_2}{g_1+g_2}.
\eea
\end{subequations}
\begin{figure}
\includegraphics[bb=50 50 554 770,angle=270,scale=0.46]{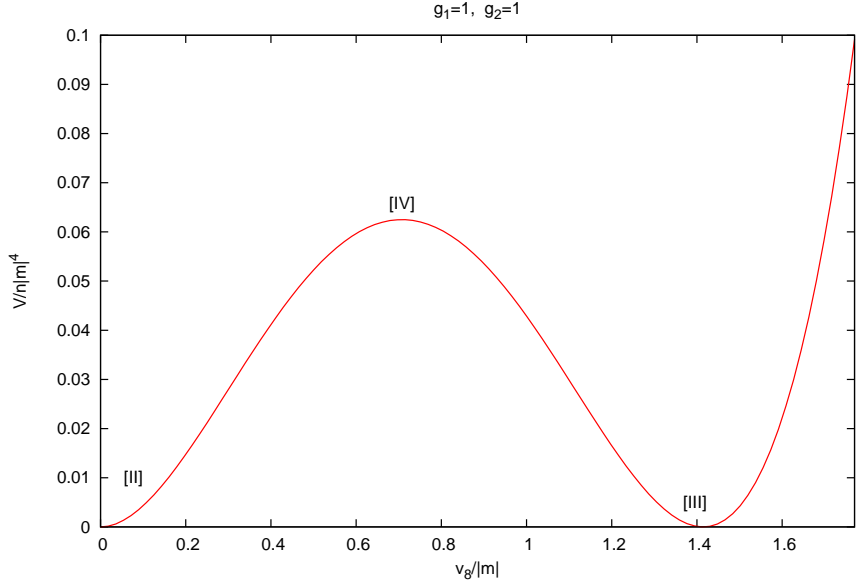}
\caption{Potential energy at $n=\infty$ as a function of $v_8$ after rescaling $v_8\longrightarrow \sqrt{n}v_8$. The value of $v_0$ is set to where local minima were found. The energy is shifted by a constant to take its zero at $v_8=0$. Surprisingly, the two minima describe the same symmetry breaking pattern.}
\end{figure}
It is interesting to note that only excitation multiplicities depend on $n$, masses do not. We see that (\ref{Eq:massIII}) describes the usual $U_L(n)\times U_R(n) \longrightarrow U_V(n)$ breaking, but the spectrum belonging to (\ref{Eq:massII}) is the same; we just have to interchange $K \longleftrightarrow \kappa$. Obtaining a massive pseudoscalar and massless scalar hints that [II] might not have physical relevance when quark masses are nonzero. We will come to this point at the end of Sec. IV. Nevertheless, since energies and spectra of these states are the same, both describe the very same symmetry breaking. Therefore, we can conclude that, at least at classical level we found two physically equivalent states, and it cannot be distinguished, which one is the ground state. At this point, we would like to emphasize that, in the large-$n$ limit it can be very misleading that extrema [II]-[III]-[IV] lie on the same line of $v_0/n=\sqrt{-m^2/(g_1+g_2)}$ and while minimum [II] has $v_8=0$, [III] shows $v_8 \neq 0$ (Fig. 2). Since the value of $v_0$ is the same for [II] and [III], one would expect that obviously these correspond to different symmetry breaking patterns, but eventually it turns out not to be true.

\section{Quantum level}
We would like to establish connections with previous works; therefore, we choose to include fluctuations in the framework of the FRG formalism \cite{wetterich93}. In FRG, one derives a flow equation for a one parameter ($k$) quantum effective action $\Gamma_k$, which interpolates between the classical action (describing microphysics) and the quantum effective action including fluctuations. For some general bosonic field(s) $\Phi$, the one parameter Schwinger functional in Euclidean space is defined as
\bea
e^{-W_k[J]}=\int {\cal D}\Phi e^{-S[\Phi]-\frac12\int\Phi\cdot R_k \cdot \Phi - \int J\cdot\Phi},
\eea
where $J$ is a source and $R_k$ is a regulator function suppressing fluctuations with momenta lower than $k$. The $k$-dependent quantum effective action is defined as
\bea
\Gamma_k[\bar{\Phi}]=W_k[J]-\int J\cdot \bar{\Phi}-\frac12\int \bar{\Phi}\cdot R_k\cdot\bar{\Phi},
\eea
which obeys the so-called flow equation:
\bea
\label{Eq:flow}
\frac{\partial \Gamma_k}{\partial k}=\frac12 \Tr \left[\frac{\partial R_k}{\partial k}(\Gamma_k^{(2)}+R_k)^{-1}\right],
\eea
where the trace has to be taken in both functional and matrix sense, and $\Gamma_k^{(2)}$ is the second functional derivative of $\Gamma_k$. The flow starts from a UV cutoff scale $\Lambda$, where by definition ($S$ being the classical action)
\bea
\Gamma_{k=\Lambda}=S,
\eea
and goes to zero including all the fluctuations, eventually reaching the quantum effective action $\Gamma_{k=0}\equiv \Gamma$.

A common approximation scheme is the LPA, where it is assumed that $\Gamma_k$ has the following form:
\bea
\Gamma_k[\Phi]=\int_x \left(\partial_i \Phi(x)\partial_i \Phi(x) + U_k(x;\Phi)\right),
\eea
where $U_k$ is the $k$-dependent effective potential. This is expected to be a good approximation if the anomalous dimension remains small. With the choice of the commonly used regulator function introduced by Litim \cite{litim01}: $R_k(q)=(k^2-q^2)\Theta(k^2-q^2)$, from (\ref{Eq:flow}), in 3+1 dimensions we get
\bea
\label{Eq:LPA}
\frac{\partial U_k}{\partial k}=\frac{k^4}{6\pi^2} \sum_i \frac{1}{k^2+M_i^2},
\eea
where $M_i$ are the eigenvalues of the mass matrix $\partial^2 U_k/\partial \Phi_i\partial \Phi_j$.

\subsection{Analysis of the $n=2$ case}
We start to analyze the quantum effects in LPA at $n=2$, which is the simplest case. Since the $U(2)$ group has only two independent invariant tensors (i.e. the previously introduced $I_1$ and $I_2$), the $k$-dependent effective potential is a two-variable function: $U_k=U_k(I_1,I_2)$. Using Appendix A, the mass matrices can be easily built up. In general we have:
\begin{subequations}
\bea
M^{2(s)}_{ij}&=&\partial_1 U_k \frac{\partial^2 I_1}{\partial s^i \partial s^j}+\partial_2 U_k \frac{\partial^2 I_2}{\partial s^i\partial s^j}+\partial_{11} U_k\frac{\partial I_1}{\partial s^i}\frac{\partial I_1}{\partial s^j}+\partial_{22} U_k\frac{\partial I_2}{\partial s^i}\frac{\partial I_2}{\partial s^j}+ \partial_{12}U_k\left[\frac{\partial I_1}{\partial s^i}\frac{\partial I_2}{\partial s^j}+\frac{\partial I_1}{\partial s^j}\frac{\partial I_2}{\partial s^i}\right], \\
M^{2(\pi)}_{ij}&=&\partial_1 U_k \frac{\partial^2 I_1}{\partial \pi^i \partial \pi^j}+\partial_2 U_k \frac{\partial^2 I_2}{\partial \pi^i\partial \pi^j}+\partial_{11} U_k\frac{\partial I_1}{\partial \pi^i}\frac{\partial I_1}{\partial \pi^j}+\partial_{22} U_k\frac{\partial I_2}{\partial \pi^i}\frac{\partial I_2}{\partial \pi^j}+ \partial_{12}U_k\left[\frac{\partial I_1}{\partial \pi^i}\frac{\partial I_2}{\partial \pi^j}+\frac{\partial I_1}{\partial \pi^j}\frac{\partial I_2}{\partial \pi^i}\right].
\eea
\end{subequations}
With the help of (\ref{Eq:deriv1}) and (\ref{Eq:deriv2}), we can evaluate these matrices at $(v_0, v_3)$, and after diagonalization in the $0-3$ sectors, we recover the results of \cite{patkos12}: the right-hand side of (\ref{Eq:LPA}) indeed depends only on $I_1$ and $I_2$, as it should. The most important observation is that $I_1$ and $I_2$ are invariant under the transformation $v_0 \longleftrightarrow v_3$, therefore so does the right-hand side of (\ref{Eq:LPA}). If we start to flow the effective potential, this leads to the observation that since $U_{k=\Lambda}$ itself has $v_0 \longleftrightarrow v_3$ symmetry, it is preserved throughout the flow to $k=0$. In other words, on the two-dimensional plane ($v_0, v_3$), the line $v_3=0$ can be exactly mapped onto the line $v_0=0$, which leads us to the fact that the two local minima of (\ref{Eq:eosfull}) evolve along these lines exactly the same way, being undistinguishable physically even at the very end of the flow (i.e. the quantum effective action). This concludes that even though we observed a potential symmetry breaking scenario $U_L(n)\times U_R(n) \longrightarrow U_V(n-1)$, at least at $n=2$ this turns out to be $U_L(n)\times U_R(n) \longrightarrow U_V(n)$ even at quantum level. Note that, in \cite{berges97,berges97b,fukushima11} it is shown numerically that, the system undergoes a first order transition starting from the minimum of the line $v_3=0$; therefore, it does so from the minimum of the line $v_0=0$.

Since we used LPA, and restricted ourselves to the $n=2$ case, we have not yet reached our final goal: we need to analyze the situation carefully for $n>2$ without having any approximation. The previous analysis however can easily be generalized, leading us to an exact result.

\subsection{Analysis of the $n\geq 2$ case}

In this subsection we present a method which is not based on any approximation, and can be applied for arbitrary $n$. We start from the exact flow equation (\ref{Eq:flow}). The right-hand side is $U_L(n)\times U_R(n)$ invariant by construction, but this time we have more than two invariants, which can be introduced by fluctuations. Since we are interested in the effective potential at homogeneous field configurations, we do not have to care about invariant terms containing field derivatives. The terms which are to appear and relevant to our interest in the effective action are the following traces:
\bea
\Tr(MM^\dagger), \qquad \Tr(MM^\dagger MM^\dagger), \qquad \Tr(MM^\dagger MM^\dagger MM^\dagger), \qquad ...
\eea
Depending on the value $n$, we have various numbers of independent invariants evolving throughout the flow. The important observation is that all of these are made out of traces of products of $M^\dagger M$. Let us take a closer look at this entity in a background $v_0, v_8$. Since generators $T^0$ and $T^8$ are real, $M|_{v_0,v_8}=M^\dagger|_{v_0,v_8}$, therefore
\bea
MM^\dagger|_{v_0,v_8}&=&(v_0T^0+v_8T^8)^2=\left[\sqrt{\frac{1}{2n}}
\begin{pmatrix}
1 & 0 & . & . & . \\
0 & 1 & & \\
. & & 1 & \\
. & & & . & \\
. & & & & 1 \\
\end{pmatrix}
v_0+\sqrt{\frac{1}{2n(n-1)}}
\begin{pmatrix}
1 & 0 & . & . & . \\
0 & 1 & & \\
. & & 1 & \\
. & & & . & \\
. & & & & -(n-1) \\
\end{pmatrix}v_8
\right]^2 \\
&=&\frac{1}{2n}\diag\left(\left(v_0+\frac{v_8}{\sqrt{n-1}}\right)^2, \left(v_0+\frac{v_8}{\sqrt{n-1}}\right)^2, ... , \left(v_0-\sqrt{n-1}v_8\right)^2\right).
\eea
The important observation here is that, there are two choices to obtain $MM^\dagger|_{v_0,v_8} \sim {\bf 1}$, which obviously describes a symmetry breaking $U_L(n)\times U_R(n) \longrightarrow U_V(n)$:
\bea
v_0+\frac{v_8}{\sqrt{n-1}}=\pm(v_0-\sqrt{n-1}v_8).
\eea
The plus sign leads to $v_8=0$, which is the obvious choice, but the nontrivial scenario is obtained choosing the minus sign:
\bea
v_8=2\frac{\sqrt{n-1}}{n-2}v_0.
\label{Eq:minus}
\eea
(For $n=2$ we get $v_0=0$, as we should.) Checking (\ref{Eq:eosc}) (solution [III]), it lies exactly on this line. The existence of that minimum is therefore a consequence of this symmetry, which will be formulated more precisely in the next section. We see that every point of the $v_8=0$ line can be associated with a point of the line $v_8=2\frac{\sqrt{n-1}}{n-2}v_0$ in the sense that $MM^\dagger$ does not change. Because the right-hand side of (\ref{Eq:flow}) has to depend only on $MM^\dagger$ (and not $M$ itself), the aforementioned two lines evolve exactly the same during the FRG flow. Furthermore, because of continuity, not just the lines themselves, but also an open neighborhood of them have this property. This leads us to the observation that, along these lines not just the value of the potential, but also the appropriately chosen directional derivatives are the same. That is what we saw before explicitly; at classical level the calculated spectrum was the same, just some directions of the $(s^a,\pi^a)$ space had to be interchanged. This result shows that the applied heavy scalar approximation of Refs. \cite{fejos11,fejos12} might look reasonable, but obviously hinders the way in which the exact result was obtained: without having the scalar excitations included, the $U_L(n) \times U_R(n)$ invariance is broken in (\ref{Eq:flow}).

\subsection{Minima of the effective potential for a general condensate}

We can formulate the appearance of the new minimum and the symmetry transformation behind it in a more general way, and argue that, it is a discrete subset of axial symmetries which generates new minima. Let us start from a condensate proportional to the unit matrix: $<\!\!M\!\!>=v_0T^0$. Since $T^0$ commutes with everything, it is clear that $U_V(n)$ transformations leave $<\!\!M\!\!>$ invariant, while $U_A(n)$ does not:
\bea
<\!\!M\!\!> \longrightarrow V <\!\!M\!\!> V^\dagger = <\!\!M\!\!>, \hspace{1.2cm} <\!\!M\!\!> \longrightarrow A^\dagger <\!\!M\!\!> A^\dagger = <\!\!M\!\!> (A^\dagger)^2.
\eea 
The effective potential is symmetric under $U_V(n)\times U_A(n)$ rotations, but since vector symmetries leave $<\!\!M\!\!>$ invariant, we need to search for appropriate axial-symmetry transformations to get copies of the $U_V(n)$ symmetric vacuum. With the notation $U=(A^\dagger)^2$, the most general transformation we can apply is $<\!\!M\!\!>\longrightarrow <\!\!M\!\!>U$, where $U$ is an arbitrary unitary matrix. This immediately shows that not just traces of invariant tensors, but $<\!\!M\!\!><\!\!M^\dagger\!\!>$ itself has to remain unchanged, in agreement with the previous subsection. As discussed in Sec. II., without the loss of generality, we can search for copies of the $<\!\!M\!\!>$ background with the condensate being in the center of the Lie algebra. This means we search for $U$ matrices for which exist $\{\tilde{v}_a\}$ real numbers fulfilling the following equation:
\bea
<\!\!M\!\!>U=\sum_{a=\diag} \tilde{v}_aT^a.
\eea
This can be rewritten as
\bea
\label{Eq:U}
U = \sum_{a=\diag} \alpha_a T^a,
\eea
where $\alpha_a=\sqrt{2n}\tilde{v}_a/v_0$. This shows that the intersection of the center of the Lie algebra and the $U(n)$ group itself in the fundamental representation connects various nontrivial vacua of the effective potential: each of these transformations refers to a copy of the original $U_V(n)$ symmetric minimum. Unitarity implies
\bea
\label{Eq:unitarity}
{\bf 1}=UU^\dagger=\sum_{a,b=\diag}\alpha_a\alpha_b T^aT^b.
\eea
Since $T^aT^b=2(d^{abc}+if^{abc})T^c$, using that $f^{abc}$ is totally antisymmetric and $T^0=\frac{1}{\sqrt{2n}}{\bf 1}$, (\ref{Eq:unitarity}) is equivalent to the conditions
\begin{subequations}
\label{Eq:unitarity2}
\bea
\frac{n}{2}&=&\sum_{a=\diag}\alpha_a\alpha_a, \hspace{1.5cm} 0=\sum_{a,b=\diag}\alpha_a \alpha_b d^{abc},
\eea
\end{subequations}
which come from linear independency of the generators. (\ref{Eq:unitarity2}) looks simple, but from a practical point of view, it is better to express unitarity in terms of matrix elements of $UU^\dagger$. Using the definition of diagonal generators given in Appendix A, from (\ref{Eq:unitarity}) we get:
\bea
{\bf 1}=\diag\left(\left[\frac{\alpha_0}{\sqrt{2n}}+\sum_{i=1}^{n-1} \frac{\alpha_i}{\sqrt{2i(i+1)}}\right]^2,\hspace{0.1cm}.\hspace{0.05cm}.\hspace{0.05cm}.\hspace{0.1cm},\left[\frac{\alpha_0}{\sqrt{2n}}-\frac{\alpha_jj}{\sqrt{2j(j+1)}}+\sum_{i=1}^{n-j-1}\frac{\alpha_{j+i}}{\sqrt{2(j+i)(j+i+1)}}\right]^2 ,\hspace{0.1cm}.\hspace{0.05cm}.\hspace{0.05cm}.\hspace{0.1cm}\right).
\label{Eq:copies}
\eea
We just show the zeroth and the $j$th element of the diagonal ($j\in \{0...,n-1\}$). From (\ref{Eq:copies}) we get $n$ equations for $n$ variables $\{\alpha_j\}$. 

We present a possible method to solve these coupled equations. The $(j-1)$th and the $j$th equation look as
\begin{subequations}
\bea
\label{Eq:j}
\Bigg|\frac{\alpha_0}{\sqrt{2n}}-\frac{\alpha_{j-1}(j-1)}{\sqrt{2j(j-1)}}+\frac{\alpha_j}{\sqrt{2j(j+1)}}+\sum_{i=1}^{n-j-1}\frac{\alpha_{j+i}}{\sqrt{2(j+i)(j+i+1)}}\Bigg|=1\\
\label{Eq:j+1}
\Bigg|\frac{\alpha_0}{\sqrt{2n}}-\frac{\alpha_jj}{\sqrt{2j(j+1)}}+\sum_{i=1}^{n-j-1}\frac{\alpha_{j+i}}{\sqrt{2(j+i)(j+i+1)}}\Bigg|=1.
\eea
\end{subequations}
If $j\neq 0$, we can plug (\ref{Eq:j+1}) to (\ref{Eq:j}), and arrive at
\bea
\label{Eq:req}
\alpha_j\sqrt{\frac{j+1}{2j}}-\alpha_{j-1}\sqrt{\frac{j-1}{2j}}\pm 1 =\pm 1, \qquad (j\in \{1...,(n-1)\}),
\eea
which is a recursion for $\alpha_j$ and where the $\pm$ sign on the left-hand side is not arbitrary, it depends on the sign of (\ref{Eq:j+1}), i.e. the one in the $j$th equation. Because of that, it is convenient to start the recursion backwards. We have not yet used the last matrix element of (\ref{Eq:unitarity2}). Requiring it to be equal to unity, we get the starting point of the recursion:
\bea
\label{Eq:reqst}
\frac{\alpha_0}{\sqrt{2n}}-\frac{\alpha_{n-1}}{\sqrt{2n}}\sqrt{n-1}=\pm 1.
\eea
In practice we start with (\ref{Eq:reqst}), choose a sign and express $\alpha_{n-1}$ with $\alpha_0$. Then we plug the expression of $\alpha_{n-1}$ into (\ref{Eq:req}) (sign on the left-hand side has already been determined with the previous choice!) for $j=n-1$, and choose a sign again for the right-hand side. The procedure continues until $j=1$, when $\alpha_1$ is determined explicitly, and with that we get the expression of $\alpha_0$ and therefore all the $\{\alpha_i\}$ immediately.

The solution described in the previous subsection refers to a set of choices, where at each but the first step, the {\it plus} sign is chosen: it leads to
\bea
\tilde{v}_0=v_0(n-2)/n,\quad \tilde{v}_{n-1}=2v_0\sqrt{n-1}/n, \quad \tilde{v}_i=0 \quad\!\!\! (i=1...,n-2),
\eea
which is consistent with (\ref{Eq:minus}) or (\ref{Eq:eosb}) with (\ref{Eq:eosc}). [Note that previously index notation $8$ was used to $(n-1)$ regardless of the value of $n$.] Choosing plus signs only leads to the trivial
\bea
\tilde{v}_0=v_0, \quad \tilde{v}_i=0 \quad\!\!\! (i=1...,n-1)
\eea
solution.

With the procedure outlined above, since there are $2^n$ ways of choosing signs, we can find in general all the $2^n$ solutions of (\ref{Eq:copies}) very easily, which connects the usual $v_0\neq 0, v_i=0$ line with appropriately chosen directions, and which give every local minima of the potential (\ref{Eq:V}) on a diagonal background. By construction, these turn out to be $U_V(n)$ symmetric ground states, and more importantly, evolve identically with the FRG flow. This illustrates that vacua appearing at classical level cannot break the $U_V(n)$ symmetry, and because of the discussed invariance of $MM^\dagger$, this property remains true after including quantum fluctuations too. We note, however, although it is not expected, but cannot be excluded, that higher order operators may develop even more new local minima not appearing at classical level.

Finally, let us make a comparison between the Vafa-Witten theorem \cite{vafa84} and our results. The theorem claims that, in vector-type gauge theories with zero $\theta$-vacuum angles (e.g. QCD), if the fermion masses are equal and non-zero, none of the vector symmetries can be broken spontaneously. Our result mainly agrees with this, remarkably without the appearance of any gauge fields, in an effective model. Adding the following explicit symmetry breaking term to the Lagrangian ($H=h_aT^a/4$):
\bea
\label{Eq:H}
{\cal L}_h=\Tr \left(H(M+M^\dagger)\right)\equiv h_as^a,
\eea
the authors of \cite{vafa84} prove that if $H \sim {\bf 1}$, which refer to equal quark masses, pattern of the symmetry breaking is $U_L(n)\times U_R(n) \longrightarrow U_V(n)$, moreover, it claims that, if $H=0$ and because of some reason new, maybe even vector-type symmetry breaking minima existed, their energy cannot be lower than the energy of the  symmetry conserving state. Our analysis extends this argument and excludes the possibility of having a $U_V(n)$ breaking state. In the chiral limit we do have multiple ground states as \cite{vafa84} suggests it as a possibility, but always the pattern $U_L(n)\times U_R(n) \longrightarrow U_V(n)$ occurs. However, in general, explicit symmetry breaking may split these states; therefore, $H\neq 0$ can create physically inequivalent stable and/or metastable states. For example $H \sim {\bf 1}$ with $h_0>0$ would clearly select solution [II] (\ref{Eq:eosb}) as the true ground state. Phenomenology shows \cite{lenaghan00} that for QCD $h_0>0$, $h_8<0$, with $h_3\approx 0$, which makes (\ref{Eq:eosb}) even more favorable energetically. Treating the model as an effective theory of the strong interaction, this points out the correctness of searching for a spontaneous symmetry breaking in the $v_0$ condensate in the first place, but it is important to stress that the vacuum structure of the model shows a much richer structure than it has been investigated before. 

\section{Conclusions}
We investigated the possibility of having a spontaneous symmetry breaking pattern of the $U_L(n) \times U_R(n)$ meson model other than $U_V(n).$ The Vafa-Witten theorem \cite{vafa84} states that for vector-type gauge theories with vanishing $\theta$-vacuum angles and nonzero but equal fermion masses, none of the vector-like global symmetries can be broken. Since this applies to QCD and the $U_L(n)\times U_R(n)$ model is an effective theory of the strong interaction, it is common to investigate a phase transition in a condensate proportional to the unit matrix in the latter model. However, we have no reason to believe that the model obeys a theorem valid for QCD, without further analysis.

Following the route of \cite{fejos11,fejos12}, we considered a more general $<\!\!M\!\!> = v_0T^0+v_8T^8$ type of symmetry breaking, and started our study at classical level in a region of the parameter space which is consistent with QCD ($g_2>0$). We found that new symmetry breaking minima of the effective potential appears for arbitrary $n$, also reproducing the large-$n$ result of \cite{fejos11} and a solution given in \cite{paterson81,nardi11,espinosa12}. Even though the examined pattern is capable to describe the breaking of vector symmetries, it turned out that, in the observed minima of the effective potential, all of them were conserved. Furthermore, all the states in question have the very same energy and spectrum, and therefore are physically equivalent to the ``usual'' minimum of the effective potential described by the condensate $<\!\!M\!\!> \sim {\bf 1}$.

We checked if the structure remained the same after including quantum fluctuations, with the use of the functional renormalization group approach. We started with an approximate solution, the local potential approximation for $n=2$, and found that fluctuations did not change the structure had been described before. Then, generalizing the applied arguments, we showed that the results hold even without any approximations, for arbitrary flavor number $n$. The reason behind this turned out to be the existence of a discrete subset of axialsymmetries, which connects several $U_V(n)$ symmetric vacua. This subset can be formulated as the intersection of the center of the $U(n)$ Lie algebra and the group itself. Using this symmetry, we managed to show how to obtain all the possible vacua of the theory considering the most general diagonal condensate. These minima of the effective potential are $U_V(n)$ symmetric by construction, which illustrates the validity of the Vafa-Witten theorem for the $U_L(n)\times U_R(n)$ meson model.

\section*{Acknowledgements}
The author is grateful to T. Hatsuda, A. Patk\'os, and Zs. Sz\'ep for useful suggestions and their careful reading of the manuscript. The author also thanks Y. Hidaka and Y. Tanizaki for illuminating discussions, and J. R. Espinosa, C. S. Fong and E. Nardi for their remark on the stability condition. This work was supported by the Japan Society for the Promotion of Science under ID No. P11795 and by the Foreign Postdoctoral Research program of RIKEN.

\makeatletter
\@addtoreset{equation}{section}
\makeatother 

\renewcommand{\theequation}{A\arabic{equation}} 

\appendix 
\section{Properties of the U(n) group}   

The $U(n)$ group has $n$ independent diagonal and $n(n-1)$ independent nondiagonal generators $T^i$ leading to $n^2$ in total. They are traceless and normalized as $\Tr(T^aT^b)=\delta^{ab}/2$. We introduce the generalized Gell-Mann matrices as $\lambda^i=2T^i$. The diagonal ones read as
\bea
\lambda^{(0)}&\!\!\!=\!\!\!&\sqrt{\frac{2}{n}}
\left( \begin{array}{cccc}
1 & & & \\
& 1 & & \\
& & ... & \\
& & & 1 \\
\end{array} \right), 
\qquad \lambda^{(1)}=
\left( \begin{array}{cccc}
1 & & & \\
& -1 & & \\
& & 0 & \\
& & & ... \\
\end{array} \right),\quad... \quad \lambda^{(n-1)}=\sqrt{\frac{2}{n(n-1)}}
\left( \begin{array}{cccc}
1 & & & \\
& 1 & & \\
& & ... & \\
& & & -(n-1) \\
\end{array} \right).
\eea
We note that, except for the last subsection of section IV., throughout the paper conventionally ``8'' index goes to $\lambda^{(n-1)}$, independently of the actual value of the flavor number $n$:
\bea
2T^8=\lambda^8\equiv \lambda^{(n-1)}.
\eea
The nondiagonal generators form two groups, as generalizations of $(x)-$ and $(y)$-type of Pauli matrices in the following sense:
\bea
\label{app-nondiag}
\big(\lambda^{(x,jk)}\big)_{ab}=\delta_{ak}\delta_{bj}+\delta_{aj}\delta_{bk}, \qquad \big(\lambda^{(y,jk)}\big)_{ab}=i\delta_{ak}\delta_{bj}-i\delta_{aj}\delta_{bk}.
\eea
Here compact index notations $(x,jk)$ and $(y,jk)$ ($j<k$) were introduced. The structure constants of the algebra are defined through the relation
\bea
\lambda^a\lambda^b=(d_{abc}+if_{abc})\lambda^c,
\eea
where $d_{abc}$ is totally symmetric and $f_{abc}$ is totally antisymmetric in their indices. We have
\begin{subequations}
\label{app-struct}
\bea
[\lambda^a,\lambda^b]&\!\!\!=\!\!\!&2if_{abc}\lambda^c \qquad \!\!\Longrightarrow \qquad \Tr\big[[\lambda^a,\lambda^b]\lambda^c\big]=4if_{abc}, \\
\{\lambda^a,\lambda^b\}&\!\!\!=\!\!\!&2d_{abc}\lambda^c \qquad \Longrightarrow \qquad \Tr \big[\{\lambda^a,\lambda^b\}\lambda^c\big]=4d_{abc},
\eea
\end{subequations}
where $[.,.]$ and $\{.,.\}$ refer to commutation and anticommutation, respectively. Alternatively, we can write 
\bea
f_{abc}=\frac12\Im \Tr(\lambda^a\lambda^b\lambda^c), \qquad d_{abc}=\Re \Tr(\lambda^a\lambda^b\lambda^c)/2. 
\eea
The structure constants needed for obtaining the mass matrices are
\bea
\label{Eq:structcon}
d^{0ij}&=&\sqrt{\frac{2}{n}}\delta^{ij}, \qquad d^{888}=(2-n)\sqrt{\frac{2}{n(n-1)}}, \qquad d^{8ij\neq 0,8}=\begin{cases} (2-n)\sqrt{\frac{1}{2n(n-1)}}\delta^{ij} , \hspace{0.1cm} \ifff \hspace{0.1cm} i,j \in \{(x,jn),(y,jn)\} \\ \sqrt{\frac{2}{n(n-1)}}\delta^{ij}, \hspace{1.3cm} \elsee \end{cases},\nonumber\\
d^{i88}&=&\sqrt{\frac{2}{n}}\delta^{i0}+d^{888}\delta^{i8}, \qquad f^{u8v}=\sqrt{\frac{n}{2(n-1)}}\left(\delta_{u,(y,jn)}\delta_{v,(x,jn)}-\delta_{u,(x,jn)}\delta_{v,(y,jn)}\right).
\eea

\makeatletter
\@addtoreset{equation}{section}
\makeatother 

\renewcommand{\theequation}{B\arabic{equation}}

\section{Classical mass matrices} 

In this section we show the exact form of the mass matrices at classical level. First we list the first and second derivatives of the invariants in a background $v_0, v_8$ needed for the expressions of the masses. The first derivatives:
\begin{subequations}
\bea
\frac{\partial I_1}{\partial s^i}\bigg|_{v_0,v_8}&=&v_0\delta^{i0}+v_8\delta^{i8},\qquad \frac{\partial I_2}{\partial s^i}\bigg|_{v_0,v_8}=\frac{1}{n}v_0^3\delta^{i0}+\frac{3}{\sqrt{2n}}d^{i88}v_8^2v_0+\frac{3}{n}v_0^2v_8\delta^{i8}+\frac12d^{888}d^{i88}v_8^3+\frac{1}{n}v_8^3\delta^{8i}, \\
\frac{\partial I_2}{\partial \pi^i}\bigg|_{v_0,v_8}&=&0, \qquad \frac{\partial I_2}{\partial \pi^i}\bigg|_{v_0,v_8}=0.
\eea
\label{Eq:deriv1}
\end{subequations}
The nonzero second derivatives:
\begin{subequations}
\bea
\frac{\partial^2 I_1}{\partial s^i \partial s^j}\bigg|_{v_0,v_8}&=&\delta^{ij}, \quad \frac{\partial^2 I_1}{\partial \pi^i \partial \pi^j}\bigg|_{v_0,v_8}=\delta^{ij}, \\
\frac{\partial^2 I_2}{\partial s^i \partial s^j}\bigg|_{v_0,v_8}&=&\frac{1}{n}(3v_0^2+v_8^2)\delta^{ij}+3\sqrt{\frac{2}{n}}d^{8ij}v_0v_8+\frac12d^{888}d^{8ij}v_8^2+\frac{2}{n}(\delta^{i8}\delta^{j8}+\delta^{i0}\delta^{i8})v_8^2+(d^{8ii})^2\delta^{ij\neq 0,8}v_8^2, \nonumber\\
&&+d^{888}\sqrt{\frac{2}{n}}(\delta^{i0}\delta^{j8}+\delta^{j0}\delta^{i8})v_8^2+(d^{888})^2\delta^{i8}\delta^{j8}v_8^2, \\
\frac{\partial^2 I_2}{\partial \pi^i\partial \pi^j}\bigg|_{v_0,v_8}&=&\frac{1}{n}(v_0^2+v_8^2)\delta^{ij}+\frac12d^{888}d^{8ij}v_8^2+\sqrt{\frac{2}{n}}d^{8ij}v_8v_0+\sum_a (f^{ai8})^2\delta^{ij}v_8^2.
\eea
\label{Eq:deriv2}
\end{subequations}
Relations (\ref{Eq:deriv1}) are also needed for the derivation of the field equations.  Using the expressions of the structure constants (\ref{Eq:structcon}), invariants (\ref{Eq:inv}) and their derivatives (\ref{Eq:deriv1}), (\ref{Eq:deriv2}), nonzero elements of the mass matrices are as follows:
\begin{subequations}
\label{Eq:massmatrices}
\bea
M_{ij}^{2(s)}=
\begin{cases} m^2 + \frac{1}{n^2}\left(g_1(3v_0^2+v_8^2)+3g_2(v_0^2+v_8^2)\right), \hspace{3.6cm} i=0,\hspace{0.2cm} j=0 \\
\frac{1}{n^2}\left(2g_1v_0v_8+g_2(6v_0v_8-3\frac{n-2}{\sqrt{n-1}}v_8^2)\right), \hspace{3.8cm} i=0,\hspace{0.2cm} j=8 \quad \textrm{or} \quad i=8,\hspace{0.2cm} j=0 \\
m^2+\frac{1}{n^2}\left(g_1(v_0^2+3v_8^2)+3g_2(v_0^2+\frac{n^2-3n+3}{n-1}v_8^2-2\frac{n-2}{\sqrt{n-1}}v_0v_8)\right), \hspace{0.15cm} i=8,\hspace{0.2cm} j=8\\
m^2+\frac{1}{n^2}\left(g_1(v_0^2+v_8^2)+g_2(3v_0^2+\frac{n^2-3n+3}{n-1}v_8^2-3\frac{n-2}{\sqrt{n-1}}v_0v_8) \right), \hspace{0.35cm} i=j\in \{(x,jn),(y,jn)\} \\
m^2+\frac{1}{n^2}\left(g_1(v_0^2+v_8^2)+3g_2(v_0^2+v_8^2+\frac{2}{\sqrt{n-1}}v_0v_8)\right), \hspace{1.8cm} i=j, \hspace{0.2cm} \elsee.
\end{cases} \\
M_{ij}^{2(\pi)}=
\begin{cases}m^2+\frac{1}{n^2}\left(g_1+g_2\right)\left(v_0^2+v_8^2\right), \hspace{5.35cm} i=0,\hspace{0.2cm} j=0 \\
\frac{g_2}{n^2}\left(2v_0v_8-\frac{n-2}{\sqrt{n-1}}v_8^2\right), \hspace{6.23cm} i=0,\hspace{0.2cm} j=8 \quad \textrm{or} \quad i=8,\hspace{0.2cm} j=0 \\
m^2+\frac{1}{n^2}\left(g_1(v_0^2+v_8^2)+g_2(v_0^2+\frac{n^2-3n+3}{n-1}v_8^2-2\frac{n-2}{\sqrt{n-1}}v_0v_8)\right),\hspace{0.55cm} i=8,\hspace{0.2cm} j=8\\
m^2+\frac{1}{n^2}\left(g_1(v_0^2+v_8^2)+g_2(v_0^2+\frac{n^2-n+1}{n-1}v_8^2-\frac{n-2}{\sqrt{n-1}}v_0v_8) \right), \hspace{0.9cm} i=j\in \{(x,jn),(y,jn)\}\\
m^2+\frac{1}{n^2}\left(g_1(v_0^2+v_8^2)+g_2(v_0^2+\frac{1}{n-1}v_8^2+\frac{2}{\sqrt{n-1}}v_0v_8)\right),  \hspace{1.45cm}  i=j, \hspace{0.2cm} \elsee.
\end{cases}
\eea
\end{subequations}

\end{document}